\documentclass[amsmath, amssymb, english, aps, prl, twocolumn, reprint,floatfix, superscriptaddress]{revtex4-1}

\usepackage[english]{babel}
\usepackage{graphicx}
\usepackage{verbatim}
\usepackage[usenames,dvipsnames]{color}

\def\t2g{t${}_{2g}$}

\def\SRO{Sr$_2$RuO$_4$\,}
\newcommand{\TFL}{$T_{\mathrm{FL}}$\,}

\begin{document}

\author{Jernej~Mravlje}
\affiliation{Jo\v{z}ef Stefan Institute, Jamova~39, Ljubljana, Slovenia}

\author{Antoine~Georges}
\affiliation{Coll\`ege de France, 11 place Marcelin Berthelot, 75005 Paris, France}
\affiliation{Centre de Physique Th\'eorique, \'Ecole Polytechnique, CNRS,
91128 Palaiseau Cedex, France}
\affiliation{DQMP, Universit\'e de Gen\`eve, 24 quai Ernest Ansermet, CH-1211 Gen\`eve, Suisse}

\begin{abstract}
We calculate the in-plane Seebeck coefficient of \SRO within a framework combining electronic structure and dynamical mean-field theory. 
We show that its temperature-dependence is consistent with entropic considerations embodied in the Kelvin formula, and 
that it provides a meaningful probe of the crossover out of the Fermi liquid regime into an incoherent metal. 
This crossover proceeds in two stages: the entropy of spin degrees of freedom is released around 
room-temperature while orbital degrees of freedom remain quenched up to much higher temperatures. 
This is confirmed by a direct calculation of the corresponding susceptibilities, and is a hallmark of `Hund's metals'. 
 We also calculate the c-axis thermopower, and predict that it exceeds substantially 
 the in-plane one at high-temperature, a peculiar behaviour which originates from an interlayer 'hole-filtering' mechanism.
\end{abstract}


\title{Thermopower and Entropy: lessons from Sr$_2$RuO$_4$} 

\maketitle

When a thermal gradient $\nabla T$ is established in a material, an electric field 
$-\nabla V$ is also generated. This thermoelectric effect can be used for
solid-state refrigeration and waste-heat recovery. The Seebeck coefficient (or thermopower) is the 
ratio $\alpha =-\nabla V/\nabla T$, measured under the condition that no electrical current flows. 
It not only determines the suitability of the material for thermoelectric applications, but is 
also a useful fundamental characterization of its electronic state. 
In metals with strong electronic correlations, the slope of the thermopower at low temperature has been
shown to scale with the linear coefficient of the specific heat~\cite{behnia04}.  
This is an example of  often noted but poorly understood relation between the thermopower and the entropy $S$. 
A precise relation has been established only for a  free electron gas at low temperature, 
in which case, %
$\alpha=-S/ne$~\cite{behnia04} , and in the high-temperature atomic limit where the Heikes formula 
$\alpha_H=-\frac{1}{e}\left(\frac{\partial S}{\partial n}\right)_E$ applies~\cite{chaikin76,doumerc1994}.  
This formula has been used to interpret the saturation of the thermopower at 
high-temperature in several transition-metal oxides~\cite{koshibae2000,klein06,uchida11}. %
More recently, it was suggested~\cite{peterson_shastry_Kelvin_prb_2010,silk2009} that the `Kelvin formula'   
$\alpha_K=-\frac{1}{e}\left(\frac{\partial S}{\partial n}\right)_T$ 
is a good approximation to the thermopower. 

Here we consider ruthenates, especially \SRO. We show %
that the thermopower provides key insights into the degrees of freedom 
which are relevant to the physics of these materials, and clarify when and how 
entropic considerations apply.   
\SRO is arguably the cleanest and best documented among strongly 
correlated oxides, with extensive studies of its low-temperature Fermi liquid (FL) behavior~\cite{bergemann03} and unconventional
superconductivity~\cite{mackenzie03}. 
Recent  work shifted the focus to properties at higher temperature and energy. 
It was shown that the low value of the scale $T_{\mathrm{FL}}\simeq 25$~K~\cite{hussey98} below which Fermi liquid behaviour is observed, 
and the large effective mass enhancement, are due to the Hund's rule coupling~\cite{mravlje11}.  
Ruthenates belong to a broader class of compounds that notably includes iron pnictides and have been
called `Hund's metals'~\cite{werner08,haule09,yin11natmat} (for a review, see~\cite{georges13}). 
Quasiparticle excitations, as revealed by a well-defined spectral
function peak, were shown to persist well above \TFL,
and the changes in the dispersion of these resilient
quasiparticles~\cite{deng13} as a function of temperature~\cite{xu13} and 
frequency~\cite{stricker14} explain the deviations from Fermi liquid behavior. 
Overall, these  results suggest that one can characterize the crossover into the 
incoherent regime better than previously imaginable.

\begin{figure}
 \begin{center}
\includegraphics[width=1\columnwidth,keepaspectratio]{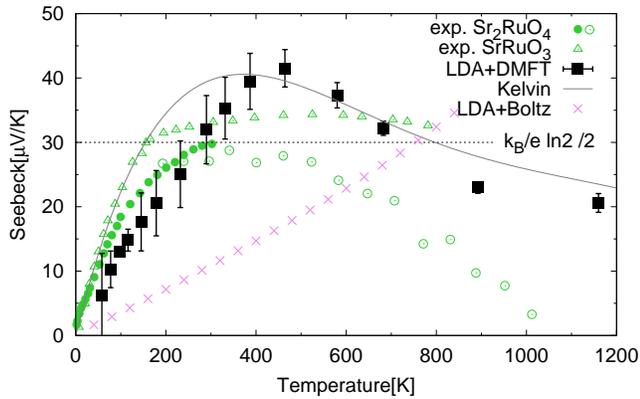}
   \end{center}
   \caption{\label{fig:seebeck}
   Temperature dependence of the in-plane thermopower in \SRO.  
    LDA+DMFT Kubo results %
    are compared to experiment at low $T$~\cite{yoshino96,xu08}  (full circles) and at high-$T$~\cite{keawprak08} (open circles), 
    and to the result of the LDA-Boltzmann theory %
     (crosses). 
     The entropic Kelvin estimate $\alpha_K=-1/e\, \partial S/\partial n|_T$ (plain line)  
     is close to the full Kubo result.  
    The dashed horizontal line indicates the Heikes-like estimate derived in the text including only spin degrees of freedom.  
    A similar behaviour is found experimentally for other ruthenates 
    (triangles: data from Ref.~\cite{klein06thesis} for SrRuO$_3$).  
    }
\end{figure}

The ab-plane thermopower of ruthenates initialy increases linearly with $T$, as expected in a FL, and saturates for 
$T\sim 300$~K at a value of order $25$-$35$~$\mu$V/K 
(see Fig.~\ref{fig:seebeck} for \SRO and SrRuO$_3$). As noted 
in Refs.~\cite{klein06,klein06thesis,hebert15}, this value depends weakly on the cation, 
the lattice structure and the doping.  
This universality is intriguing: does the value of thermopower reflect the active degrees of freedom in the metallic
state above \TFL and thereby reveal its physical nature~?
Boldly neglecting configurational and orbital contributions to the Heikes formula, 
the authors of Ref.~\cite{klein06} proposed that only spin degrees of freedom must 
be retained in the atomic entropy. 
May atomic considerations apply to an itinerant metal, and if so what happened to the 
neglected degrees of freedom ?

In this letter, we use electronic structure and dynamical mean-field theory (LDA+DMFT)  
to perform materials-realistic calculations of the thermopower of \SRO based 
on the Kubo transport formalism. 
We show that the observed value of the room-temperature thermopower is explained by 
the fact that the spin degrees of freedom are fluctuating in this range of temperature, 
while the orbital moments remain quenched up to ~$\sim 500$~K. At this temperature, 
the thermopower displays a downturn, indicating that the decoherence proceeds via 
two distinct stages, a hallmark of Hund's metals. 
We find that the in-plane thermopower is  well approximated by the Kelvin formula for all temperatures, 
establishing a connection to entropic considerations. 
We also calculate the c-axis thermopower and predict a strong enhancement at 
high temperatures, which we explain by a hole-filtering mechanism resulting from 
the crystal structure of \SRO.   

We use the LDA+DMFT method, in the implementation of Refs.~\cite{aichhorn09,aichhorn11,TRIQS},  
as applied to \SRO in earlier work~\cite{mravlje11,stricker14}. 
The rotationally invariant interaction $H=(U-3J) \hat{N}(\hat{N}-1)/2 -2J \vec{S}^2 -J \vec{L}^2/2$ was applied to the 
$t_{2g}$ atomic shell, with $\vec{S}$ and $\vec{L}$ the total spin and orbital pseudo-spin operators, respectively.  
The same interaction parameters $U=2.3$eV and $J=0.4$eV as in Refs.~\cite{mravlje11,stricker14} were used. 
The thermopower in DMFT is  
$\alpha= - k_B/e \int d\omega T(\omega) \beta \omega (-\partial f/\partial \omega)/ \int d\omega T(\omega) (-\partial f/\partial \omega)$, 
where $f(\omega)$ is the Fermi function, $\beta=1/k_B T$, and the transport function reads: 
	\begin{equation}\label{eq:seebeck}
		T(\omega)=\frac{2\pi e^2}{V}\sum_{\vec{k}}\mathrm{Tr}\left[\,v_{\vec{k}} A_{\vec{k}}^{\phantom{x}}(\omega)
		v_{\vec{k}} A_{\vec{k}}^{\phantom{x}}(\omega)\right].
	\end{equation}
Here, $V$ is a normalization volume, $A_{\vec{k}}(\omega)$ the spectral function matrix, and  
$v_{\vec{k}}$ the band velocities in the $x$-direction (in-plane response) or $z$-direction (out of plane). 
Self-energies obtained by a continuous-time Monte-Carlo impurity solver~\cite{gull11,TRIQS}
were analytically continued using both a stochastic maximum entropy method~\cite{Beach2004} and Pad\'e approximants~\cite{supp}.  
The total energies were calculated from charge self-consistent calculations as implemented in Ref.~\cite{aichhorn11}.

The calculated thermopower is shown on Fig.~\ref{fig:seebeck} (full squares with errorbars). 
It increases linearly with temperature, reaches a maximum close to 500 K and then slowly diminishes as the
temperature is further increased. 
Overall, the theoretical values agree reasonably well 
with experimental data.
 Above room-temperature, our results exceed the experimental values, even when the systematic error due to analytical
continuation is taken into account. 
(The estimation of the errorbars that are indicated on Fig.~\ref{fig:seebeck} is discussed in the supplemental material~\cite{supp} 
in which the overall consistency with Pade continuation is also presented). 
Also displayed on Fig.~\ref{fig:seebeck} is the LDA-Boltzmann estimate~\cite{madsen06}. As expected, the slope of 
$\alpha(T)$ at low-$T$ is then too small by a factor of about four, consistent with the 
mass renormalization due to correlations. 

\begin{figure}
 \begin{center}
\includegraphics[width=\columnwidth,keepaspectratio]{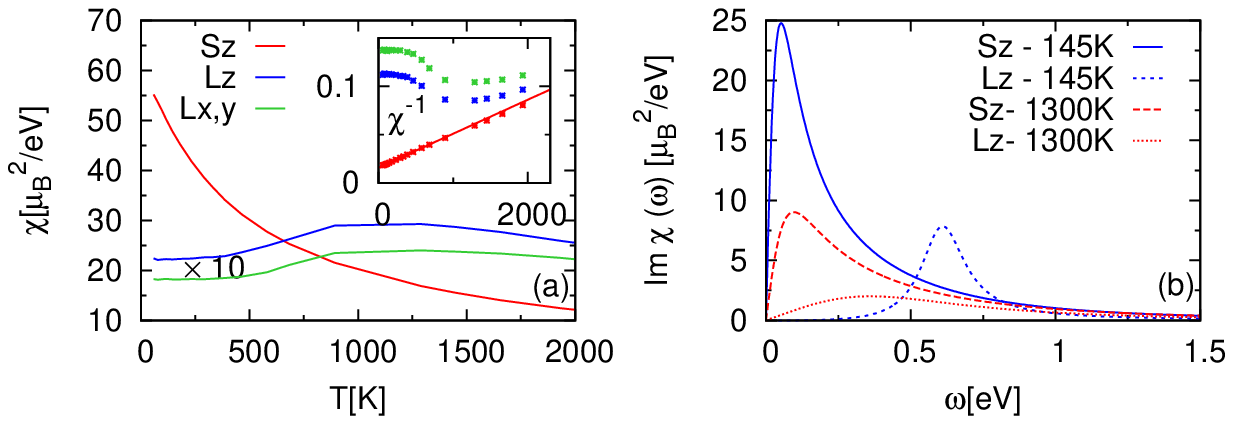}
\includegraphics[width=\columnwidth,keepaspectratio]{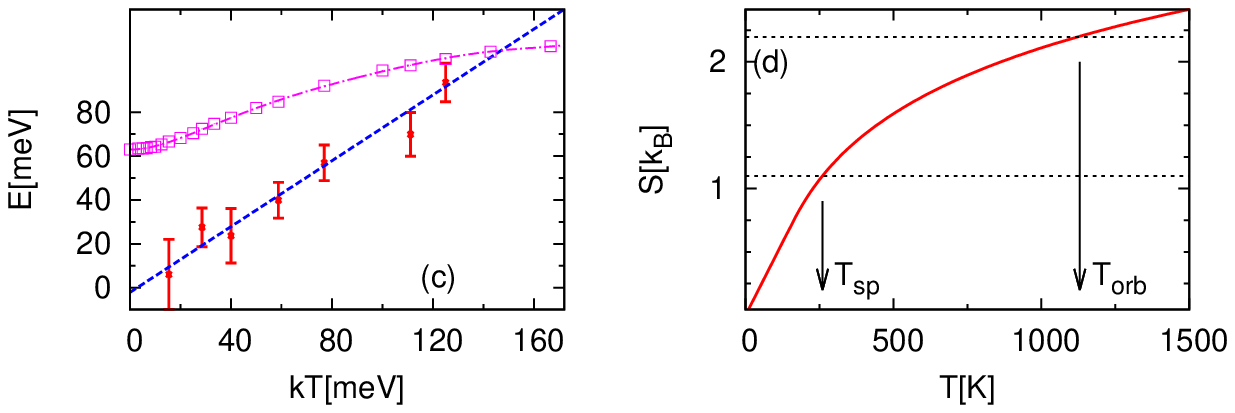}
   \end{center}
   \caption{\label{fig:chi_T}
   (a) Active degrees of freedom as revealed by  the local  
   spin and orbital susceptibilities $\chi_s,\chi_o$ (for better visibility, $10\times\chi_o$ is displayed).  
   Inset: inverse susceptibilities (symbols) - there, $\chi_o$ is multiplied by $g^2=4$
     for easier comparison with the high-$T$ local-moment behavior. 
     At intermediate $T$, the spin susceptibility can be fit well by $\chi_s = m/(T+T_m)$ (full line) with $T_m=450$~K and $m=2.5\mu_B$ 
     (close to the atomic $S=1$ moment $m=S(S+1)g^2/3\mu_B^2)$. 
     (b) Frequency-dependence  of the local spin and orbital susceptibilities. 
     (c) Total energy (circles; the dashed line is a linear fit between $200$~K and $800$~K) 
     and chemical potential (squares; the dashed-dotted line is a 6th order polynomial fit) vs. $T$.  
     (d) Estimated entropy. 
     The arrows indicate the temperatures at which the full atomic spin ($\ln 3$) and total 
     entropy (spin and orbitals, $\ln 9$) is reached. 
     }
\end{figure}

We now consider the Kelvin
approximation 
$\alpha_K=\frac{1}{e}\frac{\partial\mu}{\partial
  T}|_n=-\frac{1}{e}\frac{\partial S}{\partial n}|_T$  that
arises~\cite{peterson_shastry_Kelvin_prb_2010} if the slow
(thermodynamic) instead of the fast (transport) limit is used in the
evaluation of Onsager coefficients.  Alternatively, it can be derived
using a physical argument, requesting that the density gradient
vanishes instead of the current i.e.  $\nabla n(\mu,T)=\frac{\partial
  n}{\partial\mu}|_T\nabla\mu+\frac{\partial n}{\partial T}|_\mu\nabla
T=0$, and using the Maxwell relation $\frac{\partial n}{\partial
  T}|_\mu=-\frac{\partial n}{\partial\mu}|_T
\frac{\partial\mu}{\partial T}|_n$ into $\alpha= \nabla\mu/e\nabla T$.
Note that in the  high-$T$ limit where $\mu(T)\propto T$, the Kelvin expression $\alpha_K$ coincides 
with the Heikes formula $\alpha_H=\mu/eT$, but that it has a greater 
degree of generality and can be investigated at any temperature. 
%
%
As shown on Fig~\ref{fig:seebeck}, the Kelvin expression 
(obtained from a numerical derivative of the LDA+DMFT chemical potential~\footnote{A polynomial interpolation has been used 
to perform the numerical derivative $\partial\mu/\partial T$, see Fig.~\ref{fig:chi_T}c}) 
agrees with the Kubo result remarkably well. This finding extends to a real material the good agreement 
previously noted in model calculations~\cite{peterson_shastry_Kelvin_prb_2010,deng13,arsenault13}.  

The success of the Kelvin approximation hints at an entropic
interpretation of the thermopower. In order to identify
which degrees of freedom are active at a given temperature, 
we calculated the local spin and orbital susceptibilities, displayed on Fig.~\ref{fig:chi_T} (a,b) as 
a function of temperature and frequency. 
The spin susceptibility has a Curie-like behaviour, indicating fluctuating spin moments 
which are quenched only below the very low scale \TFL \cite{mravlje11}, which can be interpreted in DMFT  
as the Kondo temperature associated with the atomic $t_{2g}$ shell coupled to its environment.
Orbital susceptibilities behave in a drastically different manner.
They are much smaller than the magnetic one for $T\lesssim 1000$~K,
and weakly dependent on temperature up to $T\sim 500$~K. Above this
temperature, they start to increase, indicating the gradual
un-quenching of orbital degrees of freedom. At low temperature the
frequency dependence of spin- and orbital susceptibilities are
drastically different. The latter shows an activated behavior with a
peak at about $0.6$~eV, that likely comes from transitions from
$S=1,L=1$ to $S=0,L=2$ states, corresponding to an energy difference
$2J$. At high temperature, this distinction between spin and orbital
susceptibilities disappears.

We also estimated the temperature dependence of the entropy. To this
aim, we calculated the LDA+DMFT total energy as a function of $T$,
displayed in Fig.~\ref{fig:chi_T}(c) and used the thermodynamic
relation $T\partial S/\partial T=\partial E/\partial T$ (see
supplemental material \cite{supp}) to estimate the entropy.  It is
seen (Fig.~\ref{fig:chi_T}(d)) that the entropy 
corresponding to unquenched spins for Ru$^{4+}$ (spin-$1$, $S/k_B=\ln
3$) is reached by room temperature, while the entropy corresponding to
both unquenched spins and orbitals (spin-$1$, $L=1$, $S/k_B=\ln 9$) is
reached at much higher $T_\mathrm{orb}\sim 1100$~K.

These findings that reveal that the decoherence
proceeds in two stages, in which the orbital entropy is released only
after the spins are fully liberated, reinforce the Hund's metal
picture of \SRO. They are in line with the analytical study of
Refs~\cite{yin12,aron15}, in which a quantum impurity model
appropriate to these systems was considered.  This model involves
three Kondo-like coupling constants, corresponding to spin-only,
orbital-only and mixed spin-orbital degrees of freedom. It was found
that the spin coupling is much smaller than the other two, and can
even be ferromagnetic, so that the screening of orbital degrees of
freedom occurs at higher temperature, while that of the spin degrees
of freedom is controlled by the growth of the spin-orbital coupling
and occurs at a lower temperature.
Incidentally, the quenching of orbital moments at high-$T$ may explain
why calculations neglecting the spin-orbit
coupling~\cite{mravlje11,stricker14} may still be accurate for
ruthenates down to quite low $T$, even though the bare value of this
coupling is $\sim 0.1$~eV~\cite{haverkort08}. 

Having identified the relevant degrees of freedom, we can now
give an interpretation of the thermopower in terms of a simple Heikes-like 
estimate~\cite{chaikin76,doumerc1994}.  This was previously attempted
in Ref.~\cite{klein06} which considered
an intermediate valence $d^{4+x}$, mixing Ru$^{4+} (d^4)$ and Ru$^{3+} (d^5)$, for which: 
$\alpha_H=\frac{k_B}{e} \ln \frac{D(d^4)}{D(d^5)}+\frac{k_B}{e} \ln \frac{x}{1-x}$. 
The authors of  Ref.~\cite{klein06} observed that the first term gives reasonable values for the thermopower 
provided only spin degrees of freedom are retained in evaluating the degeneracies $D(d^n)$, i.e. 
$D(d^n)=2S_n+1$ with $S_n=1/2,1,3/2$ for $d^5$, $d^4$ and $d^3$, respectively  (in agreement with Hund's rule).  
The problem with this reasoning is that the second term, corresponding to configurational entropy, 
is neglected and would otherwise yield a diverging result on
approaching the actual valence $d^4$ ($x=0$) of \SRO. 
We thus reconsider the Heikes analysis  for this integer valence and obtain,     
by considering the high-$T$ limit of the chemical potential (see supplemental material for derivation~\cite{supp}): %
$\alpha_H=\mu/eT = \frac{k_B}{2e}\ln \frac{D(d^3)}{D(d^5)} = \frac{k_B}{2e} \ln 2\simeq 30\mu$V/K. 
This expression differs from the previous one on several counts. Importantly, the problematic 
configurational term does not appear. Furthermore it involves only the degeneracies of the two neighbouring 
configurations $d^3$ and $d^5$ (note also the factor $1/2$ in front of the logarithm). 
The estimated value $\sim 30\mu$V/K agrees reasonably with the observed one in the room-temperature plateau regime,  
and points to the heart of the Hund's coupling dominated nature of ruthenates:
fluctuating spins and quenched orbital moments.  Keeping orbital degrees of freedom and 
full degeneracies $D(d^3)=4$ (spin-$3/2$,$L=0$), $D(d^5)=6$ (spin $1/2$, $L=1$) would 
instead yield a negative value $k_B/2e \ln 4/6 = -17.5\mu$V/K. 
Indeed $\alpha(T)$ decreases at higher $T$, although this negative value is not
reached in our calculation perhaps because other atomic multiplets become relevant, which are 
not included in the $t_{2g}$ description. 

\begin{figure}
 \begin{center}
\includegraphics[width=\columnwidth,keepaspectratio]{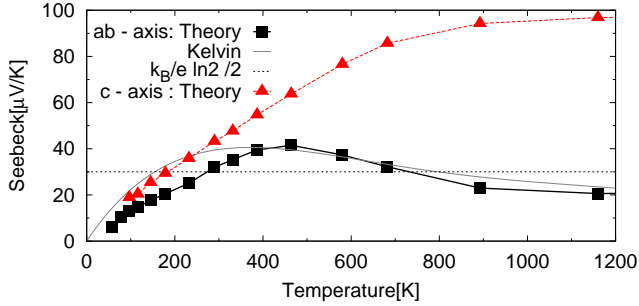}
   \end{center}
   \caption{\label{fig:seebeck_c} 
   Temperature-dependence of the calculated c-axis thermopower (triangles), 
   compared to that of the in-plane one (squares).  The errorbars (not shown) are for the c-axis response smaller than 5$\mu$V/K.
   The Kelvin and Heikes estimates are also shown, as a plain line and dotted horizontal line, respectively.
   }
\end{figure}

Although the agreement between the in-plane thermopower and the entropic 
Kelvin formula is remarkable, one has to keep in mind that 
this formula is only approximate and derived in a way that largely ignores 
the transport nature of thermoelectric coefficients. 
For instance, in a non-cubic system, these coefficients can be  
different along different crystal axis, as recently discussed for organic conductors in Ref.~\cite{kokalj14}. 
It was pointed out there (see also \cite{silk2009}) that the Kelvin formula, which includes no information 
about current matrix elements, obviously cannot account for such an anisotropy. 
We have calculated the c-axis thermopower of \SRO, and found a striking 
manifestation of this effect. As reported on Fig.~\ref{fig:seebeck_c}, the c-axis 
thermopower behaves differently from both the in-plane one and from the 
Kelvin approximation above room-temperature. It becomes remarkably large for 
a metallic system, reaching $\sim 100\mu$V/K at $T\simeq 1000$~K. 
Since, to our knowledge, the c-axis thermopower has not yet been
measured for \SRO, this finding is a prediction for future
experiments. 

To get basic insight into when the Kelvin formula can be expected to work, 
it is instructive~\cite{silk2009} to consider a non-interacting
system in which %
$\partial\mu/\partial T|_n = - k_B\int d\epsilon D(\epsilon) (\epsilon-\mu) \partial f/\partial \epsilon / 
\int d\epsilon D(\epsilon) \partial f/ \partial\epsilon$ 
with $D(\epsilon)$ the density-of-states. 
The expression of the thermopower has the same form, with the 
important difference that the transport function along the appropriate axis replaces $D(\epsilon)$.  
Hence, the Kelvin formula is expected to be a good approximation only when the transport function 
has a similar energy-dependence as the density of states. 

\begin{figure}
 \begin{center}
\includegraphics[width=1 \columnwidth,keepaspectratio]{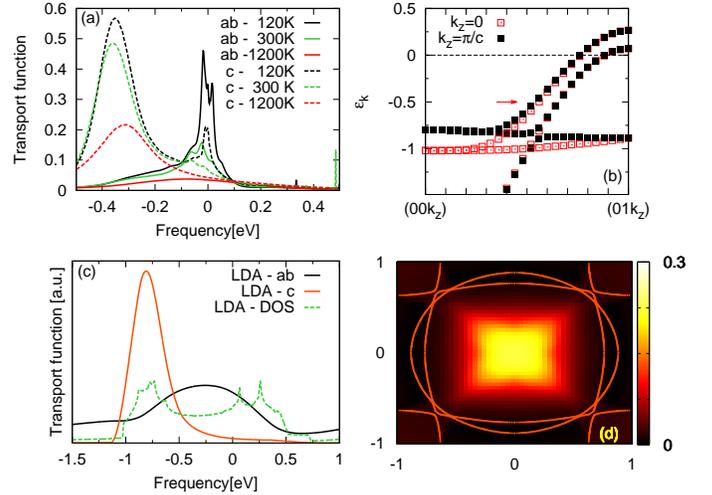}
 \end{center}
   \caption{\label{fig:transport_function} (a) Transport functions
     calculated with LDA+DMFT for different temperatures.  (b) LDA
     transport function.  (c) The band dispersion for $k_z=0$ and
     $k_z=\pi/c$ as a function of in-plane wave-vector. Note that the
     difference between the two energies $\Delta$ is maximal close to
     $k_x=k_y=0$, which corresponds to energies in the range $-0.8$ to
     $-1$~eV. This explains the strong peak observed in (c) in this
     energy range.  (d) A color intensity map of $\Delta$ in the in-plane Brillouin zone, 
     for the band denoted by an arrow in (b). The Fermi surface sheets are indicated with lines.} 
\end{figure}

In Fig.~\ref{fig:transport_function} we display the LDA+DMFT (a) and the LDA(b) transport functions. 
In contrast to the in-plane one, the c-axis transport function exhibits a pronounced 
peak at negative energies ($\sim -0.8$~eV in LDA, renormalized down to 
$\sim -0.35$eV in LDA+DMFT). 
The LDA density-of-states (dashed) has a three-peak structure, with
the outer two peaks originating from the quasi 1d xz/yz bands. The
c-axis transport function exhibits a pronounced negative frequency peak only and is thus characterized by a strong asymmetry
favouring holes. 
This occurs because, as illustrated on panels (b,c) of Fig.~\ref{fig:transport_function}, 
the dispersion of the bands as a function of $k_z$ is much stronger close to the center 
of the Brillouin zone, which corresponds to occupied states below Fermi level. 
The large c-axis thermopower is thus due to an interlayer 
`hole-filtering' mechanism.  The origin of such a behavior can be traced
to the body centered crystal structure, in which the main hopping along the
c-direction proceeds via the central atom. In a tight-binding picture,
this gives rise to a band energy of the form $\epsilon_{\vec{k}} \simeq
\epsilon(k_x,k_y)-2t_z \cos(k_k/2) \cos(k_y/2) \cos(k_z/2)$ and the
corresponding velocities in the $z-$direction are indeed maximal at
the zone center where the much larger in-plane term $\epsilon(k_x,k_y)$ 
(counted from Fermi level) is strongly negative.
The effect is largest when the thermal width of the function 
$(\epsilon-\mu)\partial f/\partial \epsilon$ becomes comparable 
to the energy of the peak in the LDA+DMFT spectral function: $5k_BT\simeq 0.35$~eV, 
i.e at temperatures $\gtrsim 800$~K. 

In summary, we have shown that the thermopower 
of ruthenates carries key physical information, and indicates in particular that the 
decoherence temperature at which the corresponding entropy is unquenched is much smaller 
for spin than for orbital degrees of freedom, a characteristic signature of a Hund's metal. 
Our calculations also predict a  different behaviour and a large value of the 
c-axis thermopower at high temperature, which results from an interlayer 
`hole-filtering' inherent to the crystal structure of \SRO. 
In contrast to the in-plane results, this illustrates the limitation of entropic 
interpretations of the thermopower. 
That entropy limitations can be overcome may be good news for applications,  
and it would be worth exploring whether such hole-filtering can be found and 
exploited in thermoelectric materials.   

\begin{acknowledgements}
We acknowledge useful discussions with K.~Behnia, R.~Daou, F.~Gascoin, S.~H\'ebert, J.~Kokalj, G.~Kotliar, A.~Maignan, S.~Shastry, L.~Taillefer,     
and the contribution of Xiaoyu Deng in the development of the transport code. 
This work was supported by the Slovenian Research Agency (ARRS) under Program P1-0044,  
by the European Research Council (ERC-319286 QMAC), and by the Swiss National Science Foundation (NCCR MARVEL). 
\end{acknowledgements}

\bibliography{thermopower}

\onecolumngrid
\newpage
\begin{center}

{\large\textbf{\boldmath
Supplemental Material\\ [0.5em] {\small to} \\ [0.5em]
Thermopower and entropy: lessons from Sr$_2$RuO$_4$}}\\[1.5em]

J. Mravlje$^1$ and A. Georges,$^{2,3,4}$\\[0.5em]

\textit{\small
$^1$Jo\v{z}ef Stefan Institute, Jamova 39, 1000 Ljubljana, Slovenia\\
$^2$Coll{\`e}ge de France, 11 place Marcelin Berthelot, 75005 Paris, France\\
$^3$Centre de Physique Th{\'e}orique, {\'E}cole Polytechnique, CNRS, 91128
Palaiseau, France \\
$^4$DQMP, Universit\'e de Gen\`eve, 24 quai Ernest Ansermet, CH-1211 Gen\`eve, Suisse\\
}

\vspace{2em}
\end{center}

\twocolumngrid
\setcounter{figure}{0}
\renewcommand{\thefigure}{SM\arabic{figure}}

\title{Supplementary Material \\ for \\ Thermopower and Entropy: lessons from Sr$_2$RuO$_4$}

\section{Generalized Heikes formula for integer occupancies in multiorbital systems}

The Heikes formula~\cite{chaikin76} approximates the Seebeck coefficient $\alpha$ as 
\begin{equation}
\alpha  \approx \alpha_H = \frac{1}{e} \left(\frac{\mu}{T}\right)_{\mathrm{at}} 
\end{equation}
In this expression, $\mu$ is the chemical potential and $()_\mathrm{at}$ indicates that $\mu/T$ is to be evaluated in the atomic
limit. 
It is actually simpler to consider $\mu/T$directly (instead of evaluating it via the thermodynamic 
relation $\mu/T=-(\partial S/\partial n)_E$ and evaluating the entropy from the number of possible
 configurations, as originally done in Ref.~\cite{chaikin76}). 

If the average occupancy $n$ is non-integer $N<n<N+1$ (for $N$
integer), $\mu/T$ in the atomic limit is controlled by the
degeneracies of the atomic state with $N$ and $N+1$ electrons, and all other 
valence states can be neglected. The occupancy is given by:
\begin{equation}
n = \frac{N d_N e^{-\beta(E_{N}-\mu N)} +(N+1) d_{N+1} e^{-\beta(E_{N+1}-\mu(N+1))}}
{d_N  e^{-\beta(E_{N}-\mu N)} + d_{N+1} e^{-\beta(E_{N+1}-\mu(N+1))}}
\end{equation}
in which $d_N$ is the degeneracy of the atomic state with $N$ electrons and $E_N$ its energy. 
Solving this equation for $\mu$ reads:
\begin{equation}
e^{\beta\mu}\,=\,\frac{d_N(n-N)}{d_{N+1} (N+1-n)}\,e^{\beta(E_{N+1}-E_N)}
\end{equation}
From this expression, it is clear that $\beta\mu$ reaches a finite value in the high-temperature limit (hence 
that $\mu\propto T$) and that the term involving $E_{N+1}-E_N$ provides only a subleading correction. As a result, 
one obtains: 
\begin{equation}
\alpha_H = \frac{k_B}{e} \log\left[\frac{d_N(n-N)}{d_{N+1} (N+1-n)}\right].
\end{equation}
which is the Heikes formula appropriate for fractional occupancy $N<n<N+1$, as 
generalized to account for atomic degeneracies in Refs.~\cite{doumerc1994,koshibae2000}

For integer occupancy $n=N$, the above calculation must be modified since both 
neighboring valence states $N\pm 1$ must be retained, as well as of course $N$ itself. 
The expression of $n=N$ now reads: 
\begin{widetext}
\begin{equation}
N =\langle n \rangle = \frac{d_{N-1} (N-1)e^{-\beta(E_{N-1}-\mu(N-1))} +
d_N N e^{-\beta(E_{N}-\mu N)} +(N+1) d_{N+1} e^{-\beta(E_{N+1}-\mu(N+1))}}{d_{N-1}e^{-\beta(E_{N-1}-\mu(N-1))} +
d_N  e^{-\beta(E_{N}-\mu N)} + d_{N+1} e^{-\beta(E_{N+1}-\mu(N+1))}}.
\end{equation}
\end{widetext}
Solving for $\mu/T$ and neglecting as above the subdominant corrections in $1/T$ now leads to the 
simple expression:  
\begin{equation}
\alpha_H= \frac{k_B}{2e} \log \frac{d_{N-1}}{d_{N+1}}.
\end{equation}
We note that this generalization of the Heikes expression appropriate to integer 
valence $n=N$ does not involve the configurational entropy term, but only the 
degeneracies of the atomic states with neighboring valence $N\pm 1$.

\section{Analytically continued self energies and comparison to Pad\'e}

\begin{figure*}
 \begin{center}
      \includegraphics[width=\columnwidth,keepaspectratio]{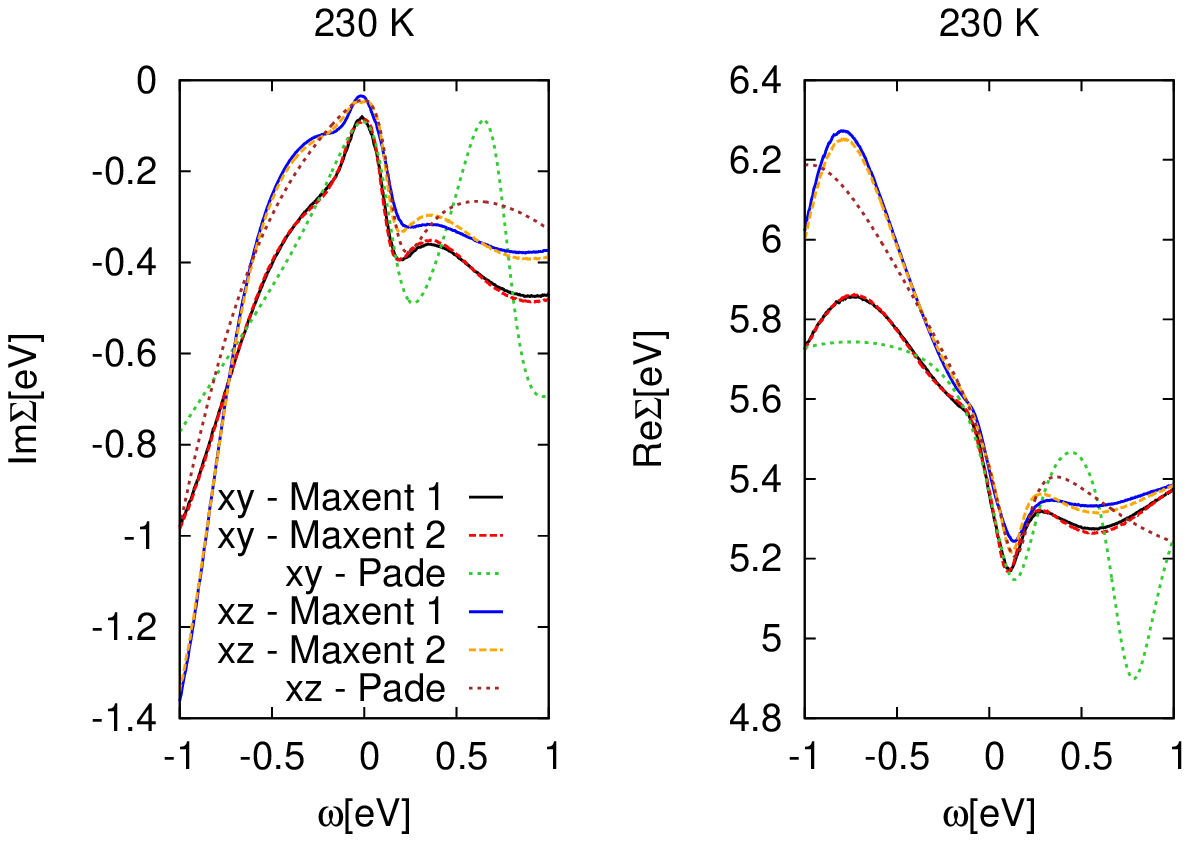} 
      \includegraphics[width=\columnwidth,keepaspectratio]{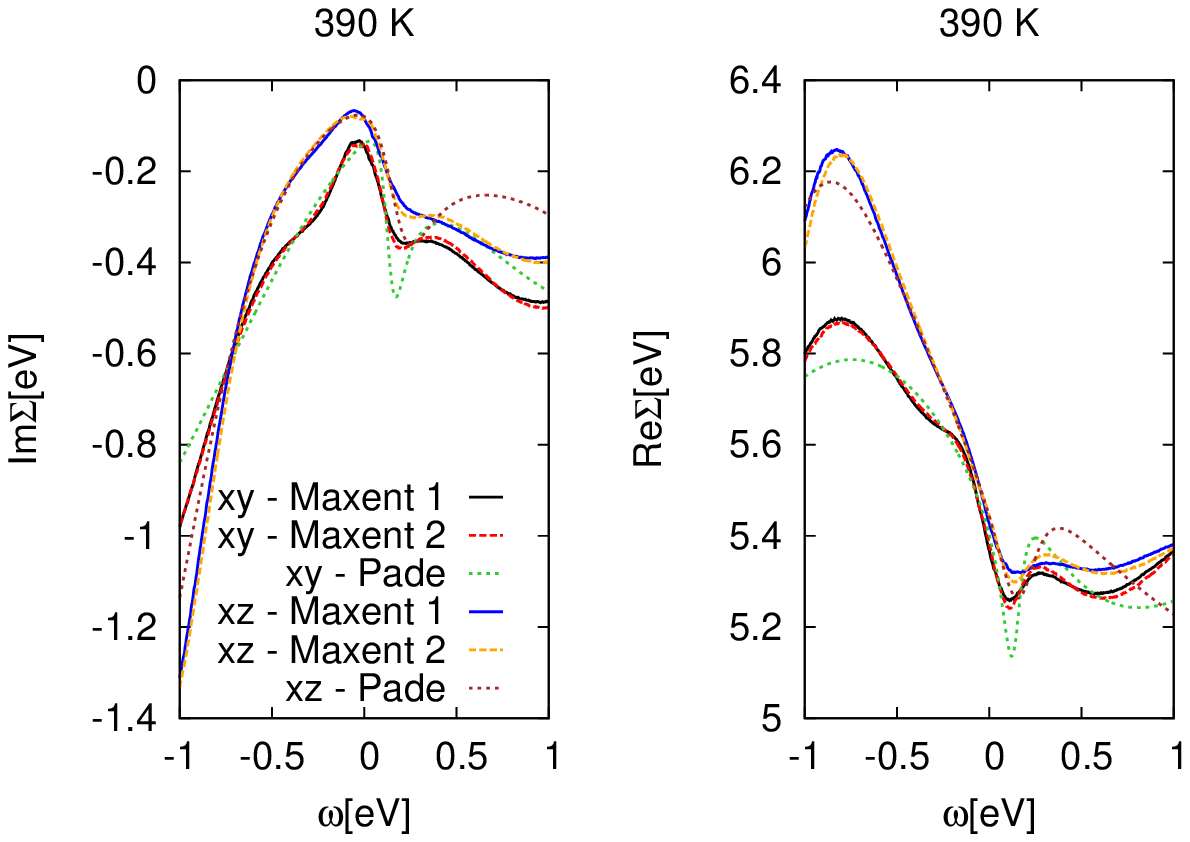} 
      \includegraphics[width=\columnwidth,keepaspectratio]{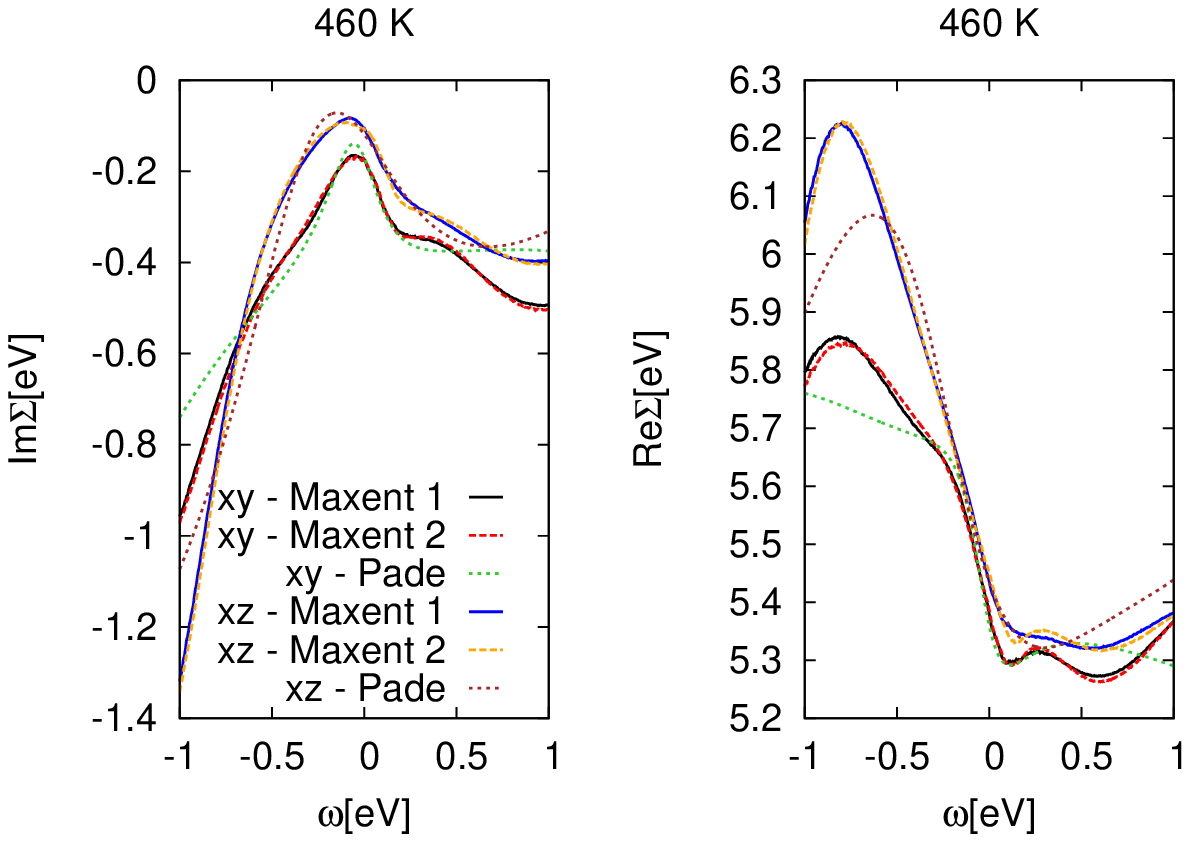} 
      \includegraphics[width=\columnwidth,keepaspectratio]{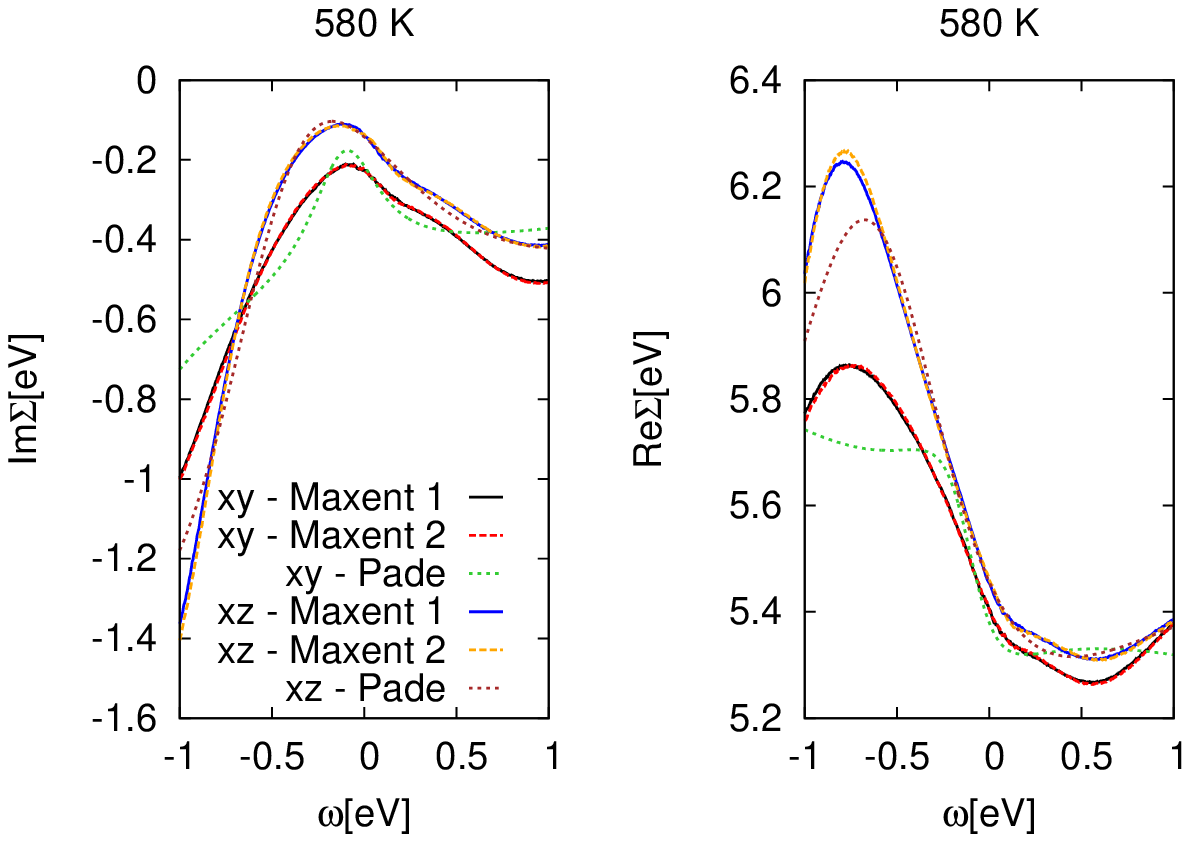} 

   \end{center}
   \caption{\label{fig:self_T}
    Analytically continued self energies. }
\end{figure*}

The analytical continuation of imaginary frequency data remains
challenging even with the most recent fast and accurate quantum impurity solvers based in continuous 
time Quantum Monte-Carlo. For the data shown
in the main text we used the self energies obtained using stochastic
Maximum Entropy approach. In practice, we analytically continue
auxiliary Green's function $\tilde{G}(i\omega) =1/(i \omega - \Sigma(i
\omega)+ \tilde{\mu})$ for a constant $\tilde{\mu}$ and extract the
self energy by inverting the resulting $\tilde{G} (\omega)$. To
estimate the error of our procedure, we set $\tilde{\mu}$ to two
different values (that should give identical answer for ideal
analytical continuation procedure on noiseless data): (1) the real
part of the Matsubara self energy at the lowest Matsubara frequency
(2) the double counting correction $(U-2J)(n-0.5)\approx 5.25$, where
$n$ is the occupancy of the $t_{2g}$ shell. The two values of
$\tilde{\mu}$ are close, the former being about 0.2eV larger than the
latter. As an independent crosscheck, we analytically continued the
self energies also directly using Pad\'e approximants. 

The real frequency self energies for selected temperatures are
presented in Fig.~\ref{fig:self_T}. One sees that the two choice of
$\tilde{\mu}$ (denoted by Maxent 1 and 2, respectively) give very similar
results. The data agrees quite well also with the self energies
obtained with Pad\'e, in particular for energy $|\omega| < 0.5eV$.

\begin{figure}
 \begin{center}
      \includegraphics[width=\columnwidth,keepaspectratio]{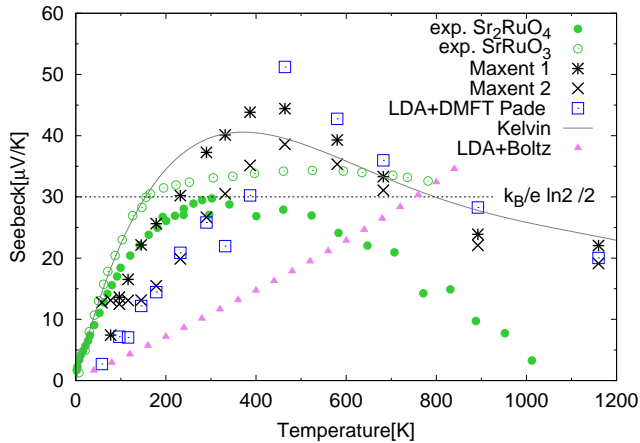}
   \end{center}
   \caption{\label{fig:pade}
Calculation of Seebeck coefficient using the Maximum entropy (stars and crosses) and Pad\'e (open squares). Other data is described in the main text. }
\end{figure}

On Fig.\ref{fig:pade} we show the calculated Seebeck coefficients
using as an input the different real frequency self energies just
discussed (stars, crosses and squares for the two choices of
$\tilde{\mu}$ and Pad\'e, respectively).  Whereas quantitatively the
differences are certainly not negligible, as the calculation of the
Seebeck coefficient is very sensitive to the low energy particle-hole
asymmetry of the imaginary part of the self energy, that is
particularly difficult to extract accurately, all the data gives
similar temperature dependence, with a maximum at about 450K.

\begin{figure}
 \begin{center}
      \includegraphics[width=\columnwidth,keepaspectratio]{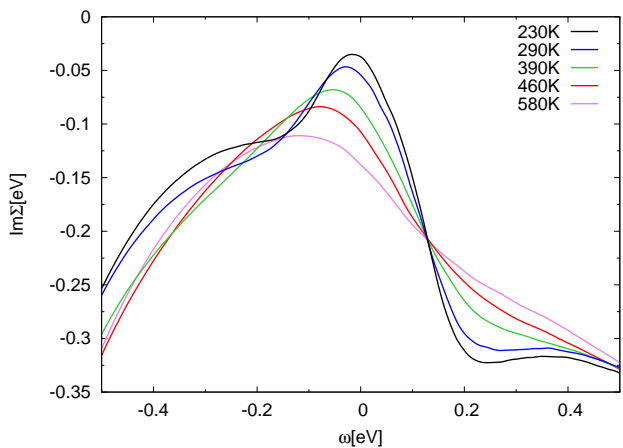}
   \end{center}
   \caption{\label{fig:self2}
Imaginary part of the self energy for the xz orbital for several temperatures. Note the progressive shift of the minimum of the $|\mathrm{Im}\Sigma|$ to the negative frequency. }
\end{figure}

Overall, the maximum entropy calculations agrees well with Pad\'e
result.  Near 450K, however, the Pad\'e data displays an abrupt
jump. The origin of this jump can be traced to abrupt shift of the
minimum of $|\mathrm{Im} \Sigma(\omega)$ towards the negative
frequency side in the Pad\'e data. The shift of the minimum that
occurs at the a temperature where the orbital moments start
unquenching (see main text), which causes the maximum in the Seebeck
coefficient occurs in the Maximum Entropy data in a milder continuous
fashion (see Fig.~\ref{fig:self2}). Pad\'e approximants are unable to
fit the self energy through the crossover, hence discontinous jump.

For the main text we used an average of the two stochastic maxent runs
in the calculation of Seebeck coefficient and the difference between
the two results was used as an indicative error bar. 

\section{Calculation of total energy and entropy}
The total energy is calculated from the charge self-consistent
LDA+DMFT calculation, as described in Ref.~\cite{aichhorn11}.  The
value of the energy and the errorbars on the plot in the main text
were obtained from the variance of the total energy in 20 consecutive
iterations of a charge self-consistent LDA+DMFT loop, starting from a
converged solution.

The precision of our data is not sufficient to obtain the entropy by direct integration of 
the thermodynamic relation
\begin{equation}
T\partial S/\partial T=\partial E/\partial T \label{eq:entropy}
\end{equation}
hence we estimate the entropy in an aproximate way.  At low-$T$, the
temperature dependence of the energy follows a Fermi-liquid form
$E=\gamma T^2/2$, with $\gamma=38$mJ/molK$^2$ for \SRO. (Mass
renormalization found in experiment is consistent with the one found
in LDA+DMFT \cite{mravlje11}).  At higher $T$, the energy is found to
depend approximately linearly on temperature.  Hence, we used a fit
$E(T)=\gamma T^2/2$ for $T<T_0$ and $E(T)=\gamma T_0^2/2+c(T-T_0)$ for
$T>T_0$, with $c=\gamma T_0$ such that $\partial E/\partial T$ is
continuous.  We find $c\simeq 0.75 k_B$ and $T_0\simeq 166$~K.  The
entropy is then obtained by integrating the thermodynamic relation
Eq.~\ref{eq:entropy}, which yields
$S=\gamma T$ for $T<T_0$ and $S=\gamma T_0 + \gamma T_0\log(T/T_0)$
for $T>T_0$. This estimate of the entropy is displayed on Fig.~2(d) of the main text.

\bibliography{thermopower}

\begin{thebibliography}{37}%
\makeatletter
\providecommand \@ifxundefined [1]{%
 \@ifx{#1\undefined}
}%
\providecommand \@ifnum [1]{%
 \ifnum #1\expandafter \@firstoftwo
 \else \expandafter \@secondoftwo
 \fi
}%
\providecommand \@ifx [1]{%
 \ifx #1\expandafter \@firstoftwo
 \else \expandafter \@secondoftwo
 \fi
}%
\providecommand \natexlab [1]{#1}%
\providecommand \enquote  [1]{``#1''}%
\providecommand \bibnamefont  [1]{#1}%
\providecommand \bibfnamefont [1]{#1}%
\providecommand \citenamefont [1]{#1}%
\providecommand \href@noop [0]{\@secondoftwo}%
\providecommand \href [0]{\begingroup \@sanitize@url \@href}%
\providecommand \@href[1]{\@@startlink{#1}\@@href}%
\providecommand \@@href[1]{\endgroup#1\@@endlink}%
\providecommand \@sanitize@url [0]{\catcode `\\12\catcode `\$12\catcode
  `\&12\catcode `\#12\catcode `\^12\catcode `\_12\catcode `\%12\relax}%
\providecommand \@@startlink[1]{}%
\providecommand \@@endlink[0]{}%
\providecommand \url  [0]{\begingroup\@sanitize@url \@url }%
\providecommand \@url [1]{\endgroup\@href {#1}{\urlprefix }}%
\providecommand \urlprefix  [0]{URL }%
\providecommand \Eprint [0]{\href }%
\providecommand \doibase [0]{http://dx.doi.org/}%
\providecommand \selectlanguage [0]{\@gobble}%
\providecommand \bibinfo  [0]{\@secondoftwo}%
\providecommand \bibfield  [0]{\@secondoftwo}%
\providecommand \translation [1]{[#1]}%
\providecommand \BibitemOpen [0]{}%
\providecommand \bibitemStop [0]{}%
\providecommand \bibitemNoStop [0]{.\EOS\space}%
\providecommand \EOS [0]{\spacefactor3000\relax}%
\providecommand \BibitemShut  [1]{\csname bibitem#1\endcsname}%
\let\auto@bib@innerbib\@empty
\bibitem [{\citenamefont {Behnia}\ \emph {et~al.}(2004)\citenamefont {Behnia},
  \citenamefont {Jaccard},\ and\ \citenamefont {Flouquet}}]{behnia04}%
  \BibitemOpen
  \bibfield  {author} {\bibinfo {author} {\bibfnamefont {K.}~\bibnamefont
  {Behnia}}, \bibinfo {author} {\bibfnamefont {D.}~\bibnamefont {Jaccard}}, \
  and\ \bibinfo {author} {\bibfnamefont {J.}~\bibnamefont {Flouquet}},\
  }\href@noop {} {\bibfield  {journal} {\bibinfo  {journal} {J. Phys.: Condens.
  Matter}\ }\textbf {\bibinfo {volume} {16}},\ \bibinfo {pages} {5187}
  (\bibinfo {year} {2004})}\BibitemShut {NoStop}%
\bibitem [{\citenamefont {Chaikin}\ and\ \citenamefont
  {Beni}(1976)}]{chaikin76}%
  \BibitemOpen
  \bibfield  {author} {\bibinfo {author} {\bibfnamefont {P.~M.}\ \bibnamefont
  {Chaikin}}\ and\ \bibinfo {author} {\bibfnamefont {G.}~\bibnamefont {Beni}},\
  }\href {\doibase 10.1103/PhysRevB.13.647} {\bibfield  {journal} {\bibinfo
  {journal} {Phys. Rev. B}\ }\textbf {\bibinfo {volume} {13}},\ \bibinfo
  {pages} {647} (\bibinfo {year} {1976})}\BibitemShut {NoStop}%
\bibitem [{\citenamefont {Doumerc}(1994)}]{doumerc1994}%
  \BibitemOpen
  \bibfield  {author} {\bibinfo {author} {\bibfnamefont {J.-P.}\ \bibnamefont
  {Doumerc}},\ }\href {\doibase http://dx.doi.org/10.1006/jssc.1994.1124}
  {\bibfield  {journal} {\bibinfo  {journal} {Journal of Solid State
  Chemistry}\ }\textbf {\bibinfo {volume} {109}},\ \bibinfo {pages} {419 }
  (\bibinfo {year} {1994})}\BibitemShut {NoStop}%
\bibitem [{\citenamefont {Koshibae}\ \emph {et~al.}(2000)\citenamefont
  {Koshibae}, \citenamefont {Tsutsui},\ and\ \citenamefont
  {Maekawa}}]{koshibae2000}%
  \BibitemOpen
  \bibfield  {author} {\bibinfo {author} {\bibfnamefont {W.}~\bibnamefont
  {Koshibae}}, \bibinfo {author} {\bibfnamefont {K.}~\bibnamefont {Tsutsui}}, \
  and\ \bibinfo {author} {\bibfnamefont {S.}~\bibnamefont {Maekawa}},\ }\href
  {\doibase 10.1103/PhysRevB.62.6869} {\bibfield  {journal} {\bibinfo
  {journal} {Phys. Rev. B}\ }\textbf {\bibinfo {volume} {62}},\ \bibinfo
  {pages} {6869} (\bibinfo {year} {2000})}\BibitemShut {NoStop}%
\bibitem [{\citenamefont {Klein}\ \emph {et~al.}(2006)\citenamefont {Klein},
  \citenamefont {H\'ebert}, \citenamefont {Maignan}, \citenamefont {Kolesnik},
  \citenamefont {Maxwell},\ and\ \citenamefont {Dabrowski}}]{klein06}%
  \BibitemOpen
  \bibfield  {author} {\bibinfo {author} {\bibfnamefont {Y.}~\bibnamefont
  {Klein}}, \bibinfo {author} {\bibfnamefont {S.}~\bibnamefont {H\'ebert}},
  \bibinfo {author} {\bibfnamefont {A.}~\bibnamefont {Maignan}}, \bibinfo
  {author} {\bibfnamefont {S.}~\bibnamefont {Kolesnik}}, \bibinfo {author}
  {\bibfnamefont {T.}~\bibnamefont {Maxwell}}, \ and\ \bibinfo {author}
  {\bibfnamefont {B.}~\bibnamefont {Dabrowski}},\ }\href {\doibase
  10.1103/PhysRevB.73.052412} {\bibfield  {journal} {\bibinfo  {journal} {Phys.
  Rev. B}\ }\textbf {\bibinfo {volume} {73}},\ \bibinfo {pages} {052412}
  (\bibinfo {year} {2006})}\BibitemShut {NoStop}%
\bibitem [{\citenamefont {Uchida}\ \emph {et~al.}(2011)\citenamefont {Uchida},
  \citenamefont {Oishi}, \citenamefont {Matsuo}, \citenamefont {Koshibae},
  \citenamefont {Onose}, \citenamefont {Mori}, \citenamefont {Fujioka},
  \citenamefont {Miyasaka}, \citenamefont {Maekawa},\ and\ \citenamefont
  {Tokura}}]{uchida11}%
  \BibitemOpen
  \bibfield  {author} {\bibinfo {author} {\bibfnamefont {M.}~\bibnamefont
  {Uchida}}, \bibinfo {author} {\bibfnamefont {K.}~\bibnamefont {Oishi}},
  \bibinfo {author} {\bibfnamefont {M.}~\bibnamefont {Matsuo}}, \bibinfo
  {author} {\bibfnamefont {W.}~\bibnamefont {Koshibae}}, \bibinfo {author}
  {\bibfnamefont {Y.}~\bibnamefont {Onose}}, \bibinfo {author} {\bibfnamefont
  {M.}~\bibnamefont {Mori}}, \bibinfo {author} {\bibfnamefont {J.}~\bibnamefont
  {Fujioka}}, \bibinfo {author} {\bibfnamefont {S.}~\bibnamefont {Miyasaka}},
  \bibinfo {author} {\bibfnamefont {S.}~\bibnamefont {Maekawa}}, \ and\
  \bibinfo {author} {\bibfnamefont {Y.}~\bibnamefont {Tokura}},\ }\href
  {\doibase 10.1103/PhysRevB.83.165127} {\bibfield  {journal} {\bibinfo
  {journal} {Phys. Rev. B}\ }\textbf {\bibinfo {volume} {83}},\ \bibinfo
  {pages} {165127} (\bibinfo {year} {2011})}\BibitemShut {NoStop}%
\bibitem [{\citenamefont {Peterson}\ and\ \citenamefont
  {Shastry}(2010)}]{peterson_shastry_Kelvin_prb_2010}%
  \BibitemOpen
  \bibfield  {author} {\bibinfo {author} {\bibfnamefont {M.~R.}\ \bibnamefont
  {Peterson}}\ and\ \bibinfo {author} {\bibfnamefont {B.~S.}\ \bibnamefont
  {Shastry}},\ }\href {\doibase 10.1103/PhysRevB.82.195105} {\bibfield
  {journal} {\bibinfo  {journal} {Phys. Rev. B}\ }\textbf {\bibinfo {volume}
  {82}},\ \bibinfo {pages} {195105} (\bibinfo {year} {2010})}\BibitemShut
  {NoStop}%
\bibitem [{\citenamefont {Silk}\ \emph {et~al.}(2009)\citenamefont {Silk},
  \citenamefont {Terasaki}, \citenamefont {Fujii},\ and\ \citenamefont
  {Schofield}}]{silk2009}%
  \BibitemOpen
  \bibfield  {author} {\bibinfo {author} {\bibfnamefont {T.~W.}\ \bibnamefont
  {Silk}}, \bibinfo {author} {\bibfnamefont {I.}~\bibnamefont {Terasaki}},
  \bibinfo {author} {\bibfnamefont {T.}~\bibnamefont {Fujii}}, \ and\ \bibinfo
  {author} {\bibfnamefont {A.~J.}\ \bibnamefont {Schofield}},\ }\href {\doibase
  10.1103/PhysRevB.79.134527} {\bibfield  {journal} {\bibinfo  {journal} {Phys.
  Rev. B}\ }\textbf {\bibinfo {volume} {79}},\ \bibinfo {pages} {134527}
  (\bibinfo {year} {2009})}\BibitemShut {NoStop}%
\bibitem [{\citenamefont {Bergemann}\ \emph {et~al.}(2003)\citenamefont
  {Bergemann} \emph {et~al.}}]{bergemann03}%
  \BibitemOpen
  \bibfield  {author} {\bibinfo {author} {\bibfnamefont {C.}~\bibnamefont
  {Bergemann}} \emph {et~al.},\ }\href@noop {} {\bibfield  {journal} {\bibinfo
  {journal} {Adv. Phys.}\ }\textbf {\bibinfo {volume} {52}},\ \bibinfo {pages}
  {639} (\bibinfo {year} {2003})}\BibitemShut {NoStop}%
\bibitem [{\citenamefont {Mackenzie}\ and\ \citenamefont
  {Maeno}(2003)}]{mackenzie03}%
  \BibitemOpen
  \bibfield  {author} {\bibinfo {author} {\bibfnamefont {A.~P.}\ \bibnamefont
  {Mackenzie}}\ and\ \bibinfo {author} {\bibfnamefont {Y.}~\bibnamefont
  {Maeno}},\ }\href {\doibase 10.1103/RevModPhys.75.657} {\bibfield  {journal}
  {\bibinfo  {journal} {Rev. Mod. Phys.}\ }\textbf {\bibinfo {volume} {75}},\
  \bibinfo {pages} {657} (\bibinfo {year} {2003})}\BibitemShut {NoStop}%
\bibitem [{\citenamefont {Hussey}\ \emph {et~al.}(1998)\citenamefont {Hussey},
  \citenamefont {Mackenzie}, \citenamefont {Cooper}, \citenamefont {Maeno},
  \citenamefont {Nishizaki},\ and\ \citenamefont {Fujita}}]{hussey98}%
  \BibitemOpen
  \bibfield  {author} {\bibinfo {author} {\bibfnamefont {N.~E.}\ \bibnamefont
  {Hussey}}, \bibinfo {author} {\bibfnamefont {A.~P.}\ \bibnamefont
  {Mackenzie}}, \bibinfo {author} {\bibfnamefont {J.~R.}\ \bibnamefont
  {Cooper}}, \bibinfo {author} {\bibfnamefont {Y.}~\bibnamefont {Maeno}},
  \bibinfo {author} {\bibfnamefont {S.}~\bibnamefont {Nishizaki}}, \ and\
  \bibinfo {author} {\bibfnamefont {T.}~\bibnamefont {Fujita}},\ }\href
  {\doibase 10.1103/PhysRevB.57.5505} {\bibfield  {journal} {\bibinfo
  {journal} {Phys. Rev. B}\ }\textbf {\bibinfo {volume} {57}},\ \bibinfo
  {pages} {5505} (\bibinfo {year} {1998})}\BibitemShut {NoStop}%
\bibitem [{\citenamefont {Mravlje}\ \emph {et~al.}(2011)\citenamefont
  {Mravlje}, \citenamefont {Aichhorn}, \citenamefont {Miyake}, \citenamefont
  {Haule}, \citenamefont {Kotliar},\ and\ \citenamefont {Georges}}]{mravlje11}%
  \BibitemOpen
  \bibfield  {author} {\bibinfo {author} {\bibfnamefont {J.}~\bibnamefont
  {Mravlje}}, \bibinfo {author} {\bibfnamefont {M.}~\bibnamefont {Aichhorn}},
  \bibinfo {author} {\bibfnamefont {T.}~\bibnamefont {Miyake}}, \bibinfo
  {author} {\bibfnamefont {K.}~\bibnamefont {Haule}}, \bibinfo {author}
  {\bibfnamefont {G.}~\bibnamefont {Kotliar}}, \ and\ \bibinfo {author}
  {\bibfnamefont {A.}~\bibnamefont {Georges}},\ }\href@noop {} {\bibfield
  {journal} {\bibinfo  {journal} {Phys. Rev. Lett.}\ }\textbf {\bibinfo
  {volume} {106}},\ \bibinfo {pages} {096401} (\bibinfo {year}
  {2011})}\BibitemShut {NoStop}%
\bibitem [{\citenamefont {Werner}\ \emph {et~al.}(2008)\citenamefont {Werner},
  \citenamefont {Gull}, \citenamefont {Troyer},\ and\ \citenamefont
  {Millis}}]{werner08}%
  \BibitemOpen
  \bibfield  {author} {\bibinfo {author} {\bibfnamefont {P.}~\bibnamefont
  {Werner}}, \bibinfo {author} {\bibfnamefont {E.}~\bibnamefont {Gull}},
  \bibinfo {author} {\bibfnamefont {M.}~\bibnamefont {Troyer}}, \ and\ \bibinfo
  {author} {\bibfnamefont {A.~J.}\ \bibnamefont {Millis}},\ }\href {\doibase
  10.1103/PhysRevLett.101.166405} {\bibfield  {journal} {\bibinfo  {journal}
  {Phys. Rev. Lett.}\ }\textbf {\bibinfo {volume} {101}},\ \bibinfo {pages}
  {166405} (\bibinfo {year} {2008})}\BibitemShut {NoStop}%
\bibitem [{\citenamefont {Haule}\ and\ \citenamefont
  {Kotliar}(2009)}]{haule09}%
  \BibitemOpen
  \bibfield  {author} {\bibinfo {author} {\bibfnamefont {K.}~\bibnamefont
  {Haule}}\ and\ \bibinfo {author} {\bibfnamefont {G.}~\bibnamefont
  {Kotliar}},\ }\href@noop {} {\bibfield  {journal} {\bibinfo  {journal} {New
  J. Phys.}\ }\textbf {\bibinfo {volume} {11}},\ \bibinfo {pages} {025021}
  (\bibinfo {year} {2009})}\BibitemShut {NoStop}%
\bibitem [{\citenamefont {Yin}\ \emph {et~al.}(2011)\citenamefont {Yin},
  \citenamefont {Haule},\ and\ \citenamefont {Kotliar}}]{yin11natmat}%
  \BibitemOpen
  \bibfield  {author} {\bibinfo {author} {\bibfnamefont {Z.~P.}\ \bibnamefont
  {Yin}}, \bibinfo {author} {\bibfnamefont {K.}~\bibnamefont {Haule}}, \ and\
  \bibinfo {author} {\bibfnamefont {G.}~\bibnamefont {Kotliar}},\ }\href@noop
  {} {\bibfield  {journal} {\bibinfo  {journal} {Nat. Mater.}\ }\textbf
  {\bibinfo {volume} {10}},\ \bibinfo {pages} {932} (\bibinfo {year}
  {2011})}\BibitemShut {NoStop}%
\bibitem [{\citenamefont {Georges}\ \emph {et~al.}(2013)\citenamefont
  {Georges}, \citenamefont {de`Medici},\ and\ \citenamefont
  {Mravlje}}]{georges13}%
  \BibitemOpen
  \bibfield  {author} {\bibinfo {author} {\bibfnamefont {A.}~\bibnamefont
  {Georges}}, \bibinfo {author} {\bibfnamefont {L.}~\bibnamefont {de`Medici}},
  \ and\ \bibinfo {author} {\bibfnamefont {J.}~\bibnamefont {Mravlje}},\
  }\href@noop {} {\bibfield  {journal} {\bibinfo  {journal} {Annu. Rev. Cond.
  Matt. Phys.}\ }\textbf {\bibinfo {volume} {4}},\ \bibinfo {pages} {137}
  (\bibinfo {year} {2013})}\BibitemShut {NoStop}%
\bibitem [{\citenamefont {Deng}\ \emph {et~al.}(2013)\citenamefont {Deng},
  \citenamefont {Mravlje}, \citenamefont {\ifmmode~\check{Z}\else
  \v{Z}\fi{}itko}, \citenamefont {Ferrero}, \citenamefont {Kotliar},\ and\
  \citenamefont {Georges}}]{deng13}%
  \BibitemOpen
  \bibfield  {author} {\bibinfo {author} {\bibfnamefont {X.}~\bibnamefont
  {Deng}}, \bibinfo {author} {\bibfnamefont {J.}~\bibnamefont {Mravlje}},
  \bibinfo {author} {\bibfnamefont {R.}~\bibnamefont {\ifmmode~\check{Z}\else
  \v{Z}\fi{}itko}}, \bibinfo {author} {\bibfnamefont {M.}~\bibnamefont
  {Ferrero}}, \bibinfo {author} {\bibfnamefont {G.}~\bibnamefont {Kotliar}}, \
  and\ \bibinfo {author} {\bibfnamefont {A.}~\bibnamefont {Georges}},\ }\href
  {\doibase 10.1103/PhysRevLett.110.086401} {\bibfield  {journal} {\bibinfo
  {journal} {Phys. Rev. Lett.}\ }\textbf {\bibinfo {volume} {110}},\ \bibinfo
  {pages} {086401} (\bibinfo {year} {2013})}\BibitemShut {NoStop}%
\bibitem [{\citenamefont {Xu}\ \emph {et~al.}(2013)\citenamefont {Xu},
  \citenamefont {Haule},\ and\ \citenamefont {Kotliar}}]{xu13}%
  \BibitemOpen
  \bibfield  {author} {\bibinfo {author} {\bibfnamefont {W.}~\bibnamefont
  {Xu}}, \bibinfo {author} {\bibfnamefont {K.}~\bibnamefont {Haule}}, \ and\
  \bibinfo {author} {\bibfnamefont {G.}~\bibnamefont {Kotliar}},\ }\href
  {\doibase 10.1103/PhysRevLett.111.036401} {\bibfield  {journal} {\bibinfo
  {journal} {Phys. Rev. Lett.}\ }\textbf {\bibinfo {volume} {111}},\ \bibinfo
  {pages} {036401} (\bibinfo {year} {2013})}\BibitemShut {NoStop}%
\bibitem [{\citenamefont {Stricker}\ \emph {et~al.}(2014)\citenamefont
  {Stricker}, \citenamefont {Mravlje}, \citenamefont {Berthod}, \citenamefont
  {Fittipaldi}, \citenamefont {Vecchione}, \citenamefont {Georges},\ and\
  \citenamefont {van~der Marel}}]{stricker14}%
  \BibitemOpen
  \bibfield  {author} {\bibinfo {author} {\bibfnamefont {D.}~\bibnamefont
  {Stricker}}, \bibinfo {author} {\bibfnamefont {J.}~\bibnamefont {Mravlje}},
  \bibinfo {author} {\bibfnamefont {C.}~\bibnamefont {Berthod}}, \bibinfo
  {author} {\bibfnamefont {R.}~\bibnamefont {Fittipaldi}}, \bibinfo {author}
  {\bibfnamefont {A.}~\bibnamefont {Vecchione}}, \bibinfo {author}
  {\bibfnamefont {A.}~\bibnamefont {Georges}}, \ and\ \bibinfo {author}
  {\bibfnamefont {D.}~\bibnamefont {van~der Marel}},\ }\href {\doibase
  10.1103/PhysRevLett.113.087404} {\bibfield  {journal} {\bibinfo  {journal}
  {Phys. Rev. Lett.}\ }\textbf {\bibinfo {volume} {113}},\ \bibinfo {pages}
  {087404} (\bibinfo {year} {2014})}\BibitemShut {NoStop}%
\bibitem [{\citenamefont {Klein}(2006)}]{klein06thesis}%
  \BibitemOpen
  \bibfield  {author} {\bibinfo {author} {\bibfnamefont {Y.}~\bibnamefont
  {Klein}},\ }\emph {\bibinfo {title} {Crucial role of the orbital and spin
  degeneracies in the thermoelectric power of metallic oxides}},\ \href@noop {}
  {Ph.D. thesis},\ \bibinfo  {school} {Universite de Caen} (\bibinfo {year}
  {2006})\BibitemShut {NoStop}%
\bibitem [{\citenamefont {H\'ebert}\ \emph {et~al.}(2015)\citenamefont
  {H\'ebert}, \citenamefont {Daou},\ and\ \citenamefont {Maignan}}]{hebert15}%
  \BibitemOpen
  \bibfield  {author} {\bibinfo {author} {\bibfnamefont {S.}~\bibnamefont
  {H\'ebert}}, \bibinfo {author} {\bibfnamefont {R.}~\bibnamefont {Daou}}, \
  and\ \bibinfo {author} {\bibfnamefont {A.}~\bibnamefont {Maignan}},\ }\href
  {\doibase 10.1103/PhysRevB.91.045106} {\bibfield  {journal} {\bibinfo
  {journal} {Phys. Rev. B}\ }\textbf {\bibinfo {volume} {91}},\ \bibinfo
  {pages} {045106} (\bibinfo {year} {2015})}\BibitemShut {NoStop}%
\bibitem [{\citenamefont {Yoshino}\ \emph {et~al.}(1996)\citenamefont
  {Yoshino}, \citenamefont {Murata}, \citenamefont {Shirakawa}, \citenamefont
  {Nishihara}, \citenamefont {Maeno},\ and\ \citenamefont
  {Fujita}}]{yoshino96}%
  \BibitemOpen
  \bibfield  {author} {\bibinfo {author} {\bibfnamefont {H.}~\bibnamefont
  {Yoshino}}, \bibinfo {author} {\bibfnamefont {K.}~\bibnamefont {Murata}},
  \bibinfo {author} {\bibfnamefont {N.}~\bibnamefont {Shirakawa}}, \bibinfo
  {author} {\bibfnamefont {Y.}~\bibnamefont {Nishihara}}, \bibinfo {author}
  {\bibfnamefont {Y.}~\bibnamefont {Maeno}}, \ and\ \bibinfo {author}
  {\bibfnamefont {T.}~\bibnamefont {Fujita}},\ }\href {\doibase
  10.1143/JPSJ.65.1548} {\bibfield  {journal} {\bibinfo  {journal} {Journal of
  the Physical Society of Japan}\ }\textbf {\bibinfo {volume} {65}},\ \bibinfo
  {pages} {1548} (\bibinfo {year} {1996})}\BibitemShut {NoStop}%
\bibitem [{\citenamefont {Xu}\ \emph {et~al.}(2008)\citenamefont {Xu},
  \citenamefont {Xu}, \citenamefont {Liu}, \citenamefont {Fobes}, \citenamefont
  {Mao}, \citenamefont {Luo},\ and\ \citenamefont {Liu}}]{xu08}%
  \BibitemOpen
  \bibfield  {author} {\bibinfo {author} {\bibfnamefont {X.~F.}\ \bibnamefont
  {Xu}}, \bibinfo {author} {\bibfnamefont {Z.~A.}\ \bibnamefont {Xu}}, \bibinfo
  {author} {\bibfnamefont {T.~J.}\ \bibnamefont {Liu}}, \bibinfo {author}
  {\bibfnamefont {D.}~\bibnamefont {Fobes}}, \bibinfo {author} {\bibfnamefont
  {Z.~Q.}\ \bibnamefont {Mao}}, \bibinfo {author} {\bibfnamefont {J.~L.}\
  \bibnamefont {Luo}}, \ and\ \bibinfo {author} {\bibfnamefont
  {Y.}~\bibnamefont {Liu}},\ }\href {\doibase 10.1103/PhysRevLett.101.057002}
  {\bibfield  {journal} {\bibinfo  {journal} {Phys. Rev. Lett.}\ }\textbf
  {\bibinfo {volume} {101}},\ \bibinfo {pages} {057002} (\bibinfo {year}
  {2008})}\BibitemShut {NoStop}%
\bibitem [{\citenamefont {Keawprak}\ \emph {et~al.}(2008)\citenamefont
  {Keawprak}, \citenamefont {Tu},\ and\ \citenamefont {Goto}}]{keawprak08}%
  \BibitemOpen
  \bibfield  {author} {\bibinfo {author} {\bibfnamefont {N.}~\bibnamefont
  {Keawprak}}, \bibinfo {author} {\bibfnamefont {R.}~\bibnamefont {Tu}}, \ and\
  \bibinfo {author} {\bibfnamefont {T.}~\bibnamefont {Goto}},\ }\href@noop {}
  {\bibfield  {journal} {\bibinfo  {journal} {Materials Transactions}\ }\textbf
  {\bibinfo {volume} {49}},\ \bibinfo {pages} {600} (\bibinfo {year}
  {2008})}\BibitemShut {NoStop}%
\bibitem [{\citenamefont {Aichhorn}\ \emph {et~al.}(2009)\citenamefont
  {Aichhorn}, \citenamefont {Pourovskii}, \citenamefont {Vildosola},
  \citenamefont {Ferrero}, \citenamefont {Parcollet}, \citenamefont {Miyake},
  \citenamefont {Georges},\ and\ \citenamefont {Biermann}}]{aichhorn09}%
  \BibitemOpen
  \bibfield  {author} {\bibinfo {author} {\bibfnamefont {M.}~\bibnamefont
  {Aichhorn}}, \bibinfo {author} {\bibfnamefont {L.}~\bibnamefont
  {Pourovskii}}, \bibinfo {author} {\bibfnamefont {V.}~\bibnamefont
  {Vildosola}}, \bibinfo {author} {\bibfnamefont {M.}~\bibnamefont {Ferrero}},
  \bibinfo {author} {\bibfnamefont {O.}~\bibnamefont {Parcollet}}, \bibinfo
  {author} {\bibfnamefont {T.}~\bibnamefont {Miyake}}, \bibinfo {author}
  {\bibfnamefont {A.}~\bibnamefont {Georges}}, \ and\ \bibinfo {author}
  {\bibfnamefont {S.}~\bibnamefont {Biermann}},\ }\href {\doibase
  10.1103/PhysRevB.80.085101} {\bibfield  {journal} {\bibinfo  {journal} {Phys.
  Rev. B}\ }\textbf {\bibinfo {volume} {80}},\ \bibinfo {pages} {085101}
  (\bibinfo {year} {2009})}\BibitemShut {NoStop}%
\bibitem [{\citenamefont {Aichhorn}\ \emph {et~al.}(2011)\citenamefont
  {Aichhorn}, \citenamefont {Pourovskii},\ and\ \citenamefont
  {Georges}}]{aichhorn11}%
  \BibitemOpen
  \bibfield  {author} {\bibinfo {author} {\bibfnamefont {M.}~\bibnamefont
  {Aichhorn}}, \bibinfo {author} {\bibfnamefont {L.}~\bibnamefont
  {Pourovskii}}, \ and\ \bibinfo {author} {\bibfnamefont {A.}~\bibnamefont
  {Georges}},\ }\href {\doibase 10.1103/PhysRevB.84.054529} {\bibfield
  {journal} {\bibinfo  {journal} {Phys. Rev. B}\ }\textbf {\bibinfo {volume}
  {84}},\ \bibinfo {pages} {054529} (\bibinfo {year} {2011})}\BibitemShut
  {NoStop}%
\bibitem [{\citenamefont {Ferrero}\ and\ \citenamefont {Parcollet}()}]{TRIQS}%
  \BibitemOpen
  \bibfield  {author} {\bibinfo {author} {\bibfnamefont {M.}~\bibnamefont
  {Ferrero}}\ and\ \bibinfo {author} {\bibfnamefont {O.}~\bibnamefont
  {Parcollet}},\ }\href@noop {} {}\bibinfo {note} {\,TRIQS: A Toolbox for
  Research on Interacting Quantum Systems,
  http://ipht.cea.fr/triqs}\BibitemShut {NoStop}%
\bibitem [{\citenamefont {Gull}\ \emph {et~al.}(2011)\citenamefont {Gull},
  \citenamefont {Millis}, \citenamefont {Lichtenstein}, \citenamefont
  {Rubtsov}, \citenamefont {Troyer},\ and\ \citenamefont {Werner}}]{gull11}%
  \BibitemOpen
  \bibfield  {author} {\bibinfo {author} {\bibfnamefont {E.}~\bibnamefont
  {Gull}}, \bibinfo {author} {\bibfnamefont {A.~J.}\ \bibnamefont {Millis}},
  \bibinfo {author} {\bibfnamefont {A.~I.}\ \bibnamefont {Lichtenstein}},
  \bibinfo {author} {\bibfnamefont {A.~N.}\ \bibnamefont {Rubtsov}}, \bibinfo
  {author} {\bibfnamefont {M.}~\bibnamefont {Troyer}}, \ and\ \bibinfo {author}
  {\bibfnamefont {P.}~\bibnamefont {Werner}},\ }\href {\doibase
  10.1103/RevModPhys.83.349} {\bibfield  {journal} {\bibinfo  {journal} {Rev.
  Mod. Phys.}\ }\textbf {\bibinfo {volume} {83}},\ \bibinfo {pages} {349}
  (\bibinfo {year} {2011})}\BibitemShut {NoStop}%
\bibitem [{\citenamefont {{Beach}}()}]{Beach2004}%
  \BibitemOpen
  \bibfield  {author} {\bibinfo {author} {\bibfnamefont {K.~S.~D.}\
  \bibnamefont {{Beach}}},\ }\href@noop {} {\ }\bibinfo {note}
  {ArXiv:cond-mat/0403055}\BibitemShut {NoStop}%
\bibitem [{sup()}]{supp}%
  \BibitemOpen
  \href@noop {} {}\bibinfo {note} {See Supplemental Material}\BibitemShut
  {NoStop}%
\bibitem [{\citenamefont {Madsen}\ and\ \citenamefont
  {Singh}(2006)}]{madsen06}%
  \BibitemOpen
  \bibfield  {author} {\bibinfo {author} {\bibfnamefont {G.}~\bibnamefont
  {Madsen}}\ and\ \bibinfo {author} {\bibfnamefont {D.}~\bibnamefont {Singh}},\
  }\href@noop {} {\bibfield  {journal} {\bibinfo  {journal} {Comp. Phys.
  Comm.}\ }\textbf {\bibinfo {volume} {175}},\ \bibinfo {pages} {6771}
  (\bibinfo {year} {2006})}\BibitemShut {NoStop}%
\bibitem [{Note1()}]{Note1}%
  \BibitemOpen
  \bibinfo {note} {A polynomial interpolation has been used to perform the
  numerical derivative $\partial \mu /\partial T$, see Fig.~\ref
  {fig:chi_T}c}\BibitemShut {NoStop}%
\bibitem [{\citenamefont {Arsenault}\ \emph {et~al.}(2013)\citenamefont
  {Arsenault}, \citenamefont {Shastry}, \citenamefont {S\'emon},\ and\
  \citenamefont {Tremblay}}]{arsenault13}%
  \BibitemOpen
  \bibfield  {author} {\bibinfo {author} {\bibfnamefont {L.-F.}\ \bibnamefont
  {Arsenault}}, \bibinfo {author} {\bibfnamefont {B.~S.}\ \bibnamefont
  {Shastry}}, \bibinfo {author} {\bibfnamefont {P.}~\bibnamefont {S\'emon}}, \
  and\ \bibinfo {author} {\bibfnamefont {A.-M.~S.}\ \bibnamefont {Tremblay}},\
  }\href {\doibase 10.1103/PhysRevB.87.035126} {\bibfield  {journal} {\bibinfo
  {journal} {Phys. Rev. B}\ }\textbf {\bibinfo {volume} {87}},\ \bibinfo
  {pages} {035126} (\bibinfo {year} {2013})}\BibitemShut {NoStop}%
\bibitem [{\citenamefont {Yin}\ \emph {et~al.}(2012)\citenamefont {Yin},
  \citenamefont {Haule},\ and\ \citenamefont {Kotliar}}]{yin12}%
  \BibitemOpen
  \bibfield  {author} {\bibinfo {author} {\bibfnamefont {Z.~P.}\ \bibnamefont
  {Yin}}, \bibinfo {author} {\bibfnamefont {K.}~\bibnamefont {Haule}}, \ and\
  \bibinfo {author} {\bibfnamefont {G.}~\bibnamefont {Kotliar}},\ }\href
  {\doibase 10.1103/PhysRevB.86.195141} {\bibfield  {journal} {\bibinfo
  {journal} {Phys. Rev. B}\ }\textbf {\bibinfo {volume} {86}},\ \bibinfo
  {pages} {195141} (\bibinfo {year} {2012})}\BibitemShut {NoStop}%
\bibitem [{\citenamefont {Aron}\ and\ \citenamefont {Kotliar}(2015)}]{aron15}%
  \BibitemOpen
  \bibfield  {author} {\bibinfo {author} {\bibfnamefont {C.}~\bibnamefont
  {Aron}}\ and\ \bibinfo {author} {\bibfnamefont {G.}~\bibnamefont {Kotliar}},\
  }\href {\doibase 10.1103/PhysRevB.91.041110} {\bibfield  {journal} {\bibinfo
  {journal} {Phys. Rev. B}\ }\textbf {\bibinfo {volume} {91}},\ \bibinfo
  {pages} {041110} (\bibinfo {year} {2015})}\BibitemShut {NoStop}%
\bibitem [{\citenamefont {Haverkort}\ \emph {et~al.}(2008)\citenamefont
  {Haverkort}, \citenamefont {Elfimov}, \citenamefont {Tjeng}, \citenamefont
  {Sawatzky},\ and\ \citenamefont {Damascelli}}]{haverkort08}%
  \BibitemOpen
  \bibfield  {author} {\bibinfo {author} {\bibfnamefont {M.~W.}\ \bibnamefont
  {Haverkort}}, \bibinfo {author} {\bibfnamefont {I.~S.}\ \bibnamefont
  {Elfimov}}, \bibinfo {author} {\bibfnamefont {L.~H.}\ \bibnamefont {Tjeng}},
  \bibinfo {author} {\bibfnamefont {G.~A.}\ \bibnamefont {Sawatzky}}, \ and\
  \bibinfo {author} {\bibfnamefont {A.}~\bibnamefont {Damascelli}},\ }\href
  {\doibase 10.1103/PhysRevLett.101.026406} {\bibfield  {journal} {\bibinfo
  {journal} {Phys. Rev. Lett.}\ }\textbf {\bibinfo {volume} {101}},\ \bibinfo
  {pages} {026406} (\bibinfo {year} {2008})}\BibitemShut {NoStop}%
\bibitem [{\citenamefont {Kokalj}\ and\ \citenamefont
  {McKenzie}(2015)}]{kokalj14}%
  \BibitemOpen
  \bibfield  {author} {\bibinfo {author} {\bibfnamefont {J.}~\bibnamefont
  {Kokalj}}\ and\ \bibinfo {author} {\bibfnamefont {R.~H.}\ \bibnamefont
  {McKenzie}},\ }\href {\doibase 10.1103/PhysRevB.91.125143} {\bibfield
  {journal} {\bibinfo  {journal} {Phys. Rev. B}\ }\textbf {\bibinfo {volume}
  {91}},\ \bibinfo {pages} {125143} (\bibinfo {year} {2015})}\BibitemShut
  {NoStop}%
\end{thebibliography}

\begin{thebibliography}{5}%
\makeatletter
\providecommand \@ifxundefined [1]{%
 \@ifx{#1\undefined}
}%
\providecommand \@ifnum [1]{%
 \ifnum #1\expandafter \@firstoftwo
 \else \expandafter \@secondoftwo
 \fi
}%
\providecommand \@ifx [1]{%
 \ifx #1\expandafter \@firstoftwo
 \else \expandafter \@secondoftwo
 \fi
}%
\providecommand \natexlab [1]{#1}%
\providecommand \enquote  [1]{``#1''}%
\providecommand \bibnamefont  [1]{#1}%
\providecommand \bibfnamefont [1]{#1}%
\providecommand \citenamefont [1]{#1}%
\providecommand \href@noop [0]{\@secondoftwo}%
\providecommand \href [0]{\begingroup \@sanitize@url \@href}%
\providecommand \@href[1]{\@@startlink{#1}\@@href}%
\providecommand \@@href[1]{\endgroup#1\@@endlink}%
\providecommand \@sanitize@url [0]{\catcode `\\12\catcode `\$12\catcode
  `\&12\catcode `\#12\catcode `\^12\catcode `\_12\catcode `\%12\relax}%
\providecommand \@@startlink[1]{}%
\providecommand \@@endlink[0]{}%
\providecommand \url  [0]{\begingroup\@sanitize@url \@url }%
\providecommand \@url [1]{\endgroup\@href {#1}{\urlprefix }}%
\providecommand \urlprefix  [0]{URL }%
\providecommand \Eprint [0]{\href }%
\providecommand \doibase [0]{http://dx.doi.org/}%
\providecommand \selectlanguage [0]{\@gobble}%
\providecommand \bibinfo  [0]{\@secondoftwo}%
\providecommand \bibfield  [0]{\@secondoftwo}%
\providecommand \translation [1]{[#1]}%
\providecommand \BibitemOpen [0]{}%
\providecommand \bibitemStop [0]{}%
\providecommand \bibitemNoStop [0]{.\EOS\space}%
\providecommand \EOS [0]{\spacefactor3000\relax}%
\providecommand \BibitemShut  [1]{\csname bibitem#1\endcsname}%
\let\auto@bib@innerbib\@empty
\bibitem [{\citenamefont {Chaikin}\ and\ \citenamefont
  {Beni}(1976)}]{chaikin76}%
  \BibitemOpen
  \bibfield  {author} {\bibinfo {author} {\bibfnamefont {P.~M.}\ \bibnamefont
  {Chaikin}}\ and\ \bibinfo {author} {\bibfnamefont {G.}~\bibnamefont {Beni}},\
  }\href {\doibase 10.1103/PhysRevB.13.647} {\bibfield  {journal} {\bibinfo
  {journal} {Phys. Rev. B}\ }\textbf {\bibinfo {volume} {13}},\ \bibinfo
  {pages} {647} (\bibinfo {year} {1976})}\BibitemShut {NoStop}%
\bibitem [{\citenamefont {Doumerc}(1994)}]{doumerc1994}%
  \BibitemOpen
  \bibfield  {author} {\bibinfo {author} {\bibfnamefont {J.-P.}\ \bibnamefont
  {Doumerc}},\ }\href {\doibase http://dx.doi.org/10.1006/jssc.1994.1124}
  {\bibfield  {journal} {\bibinfo  {journal} {Journal of Solid State
  Chemistry}\ }\textbf {\bibinfo {volume} {109}},\ \bibinfo {pages} {419 }
  (\bibinfo {year} {1994})}\BibitemShut {NoStop}%
\bibitem [{\citenamefont {Koshibae}\ \emph {et~al.}(2000)\citenamefont
  {Koshibae}, \citenamefont {Tsutsui},\ and\ \citenamefont
  {Maekawa}}]{koshibae2000}%
  \BibitemOpen
  \bibfield  {author} {\bibinfo {author} {\bibfnamefont {W.}~\bibnamefont
  {Koshibae}}, \bibinfo {author} {\bibfnamefont {K.}~\bibnamefont {Tsutsui}}, \
  and\ \bibinfo {author} {\bibfnamefont {S.}~\bibnamefont {Maekawa}},\ }\href
  {\doibase 10.1103/PhysRevB.62.6869} {\bibfield  {journal} {\bibinfo
  {journal} {Phys. Rev. B}\ }\textbf {\bibinfo {volume} {62}},\ \bibinfo
  {pages} {6869} (\bibinfo {year} {2000})}\BibitemShut {NoStop}%
\bibitem [{\citenamefont {Aichhorn}\ \emph {et~al.}(2011)\citenamefont
  {Aichhorn}, \citenamefont {Pourovskii},\ and\ \citenamefont
  {Georges}}]{aichhorn11}%
  \BibitemOpen
  \bibfield  {author} {\bibinfo {author} {\bibfnamefont {M.}~\bibnamefont
  {Aichhorn}}, \bibinfo {author} {\bibfnamefont {L.}~\bibnamefont
  {Pourovskii}}, \ and\ \bibinfo {author} {\bibfnamefont {A.}~\bibnamefont
  {Georges}},\ }\href {\doibase 10.1103/PhysRevB.84.054529} {\bibfield
  {journal} {\bibinfo  {journal} {Phys. Rev. B}\ }\textbf {\bibinfo {volume}
  {84}},\ \bibinfo {pages} {054529} (\bibinfo {year} {2011})}\BibitemShut
  {NoStop}%
\bibitem [{\citenamefont {Mravlje}\ \emph {et~al.}(2011)\citenamefont
  {Mravlje}, \citenamefont {Aichhorn}, \citenamefont {Miyake}, \citenamefont
  {Haule}, \citenamefont {Kotliar},\ and\ \citenamefont {Georges}}]{mravlje11}%
  \BibitemOpen
  \bibfield  {author} {\bibinfo {author} {\bibfnamefont {J.}~\bibnamefont
  {Mravlje}}, \bibinfo {author} {\bibfnamefont {M.}~\bibnamefont {Aichhorn}},
  \bibinfo {author} {\bibfnamefont {T.}~\bibnamefont {Miyake}}, \bibinfo
  {author} {\bibfnamefont {K.}~\bibnamefont {Haule}}, \bibinfo {author}
  {\bibfnamefont {G.}~\bibnamefont {Kotliar}}, \ and\ \bibinfo {author}
  {\bibfnamefont {A.}~\bibnamefont {Georges}},\ }\href@noop {} {\bibfield
  {journal} {\bibinfo  {journal} {Phys. Rev. Lett.}\ }\textbf {\bibinfo
  {volume} {106}},\ \bibinfo {pages} {096401} (\bibinfo {year}
  {2011})}\BibitemShut {NoStop}%
\end{thebibliography}
\end{document}